\documentclass[aps,prd,reprint,showpacs,superscriptaddress,floatfix]{revtex4-1}
\usepackage{booktabs,amsmath}
\usepackage{amssymb}
\usepackage{array}
\usepackage{mathtools}
\usepackage{psfrag}

\usepackage{amsthm}
\usepackage{newlfont}
\usepackage{esdiff}
\usepackage{multirow}
\usepackage{graphics}
\usepackage{graphicx}
\usepackage{ae,aecompl,color}

\usepackage{bm}
\usepackage[bookmarks=true,bookmarksopen=true,bookmarksnumbered=true,bookmarksopenlevel=3]{hyperref}

\begin{document}
\newcommand{\bd}{\begin{document}}
\newcommand{\ed}{\end{document}}
\newcommand{\bc}{\begin{center}}
\newcommand{\ec}{\end{center}}
\newcommand{\bfr}{\begin{flushright}}
\newcommand{\efr}{\end{flushright}}
\newcommand{\lt}{\left}
\newcommand{\rt}{\right}
\newcommand{\vs}{\vspace}
\newcommand{\hs}{\hspace}
\newcommand{\beq}{\begin{equation}}
\newcommand{\eeq}{\end{equation}}
\newcommand{\lb}{\linebreak}
\newcommand{\pb}{\pagebreak}
\newcommand{\mb}{\makebox}
\newcommand{\fb}{\framebox}
\newcommand{\mc}{\multicolumn}
\newcommand{\ben}{\begin{enumerate}}
\newcommand{\een}{\end{enumerate}}
\newcommand{\bit}{\begin{itemize}}
\newcommand{\eit}{\end{itemize}}
\newcommand{\un}{\underline}
\newcommand{\lefq}{\lefteqn}
\newcommand{\ba}{\begin{array}}
\newcommand{\ea}{\end{array}}
\newcommand{\beqa}{\begin{eqnarray}}
\newcommand{\eeqa}{\end{eqnarray}}
\newcommand{\beqas}{\begin{eqnarray*}}
\newcommand{\eeqas}{\end{eqnarray*}}
\newcommand{\bfg}{\begin{figure}}
\newcommand{\efg}{\end{figure}}
\newcommand{\bds}{\begin{displaymath}}
\newcommand{\eds}{\end{displaymath}}
\newcommand{\btb}{\begin{tabbing}}
\newcommand{\etb}{\end{tabbing}}
\newcommand{\para}{\parallel}
\newcommand{\pad}{\partial}
\newcommand{\nn}{\nonumber}
\newcommand{\la}{\leftarrow}
\newcommand{\ra}{\rightarrow}
\newcommand{\lgla}{\longleftarrow}
\newcommand{\lgra}{\longrightarrow}
\newcommand{\La}{\Leftarrow}\newcommand{\Ra}{\Rightarrow}
\newcommand{\Lra}{\Leftrightarrow}
\newcommand{\Lgla}{\Longleftarrow}
\newcommand{\Lgra}{\Longrightarrow}
\newcommand{\lan}{\langle}
\newcommand{\ran}{\rangle}
\renewcommand{\a}{\alpha}
\renewcommand{\b}{\beta}
\newcommand{\g}{\gamma}
\newcommand{\G}{\Gamma}
\renewcommand{\d}{\delta}
\newcommand{\eps}{\epsilon}
\newcommand{\Th}{\Theta}
\newcommand{\s}{\sigma}
\newcommand{\lam}{\lambda}
\newcommand{\D}{\Delta}
\newcommand{\vare}{\varepsilon}
\newcommand{\pr}{\prime}
\newcommand{\ro}{\rho}
\newcommand{\nab}{\nabla}
\newcommand{\m}{\mu}
\newcommand{\n}{\nu}
\newcommand{\Sg}{\Sigma}
\newcommand{\p}{\pi}
\newcommand{\R}{I\!\!R}
\newcommand{\om}{\omega}
\newcommand{\Om}{\Omega}
\newcommand{\ze}{\zeta}
\newcommand{\vart}{\vartheta}
\newcommand{\tri}{\triangle}
\newcommand{\f}{\frac}
\newcommand{\iny}{\infty}
\newcommand{\pro}{\propto}
\renewcommand{\arraystretch}{1.25}
\title{Quantum phase transitions  of the Dirac oscillator in a minimal length  scenario} 
%
%
\author{\textsc{L.~Menculini}}
\affiliation{Dipartimento di Fisica, Universit\`a degli Studi di Perugia, Via A.~Pascoli, I-06123 Perugia, Italy}
\author{\textsc{O.~Panella}}
\affiliation{Istituto Nazionale di Fisica Nucleare, Sezione di Perugia, Via A.~Pascoli, I-06123 Perugia, Italy}
\email[({\bf Corresponding Author})\\ Email: ]{orlando.panella@pg.infn.it }

\author{\textsc{P.~Roy}}
\affiliation{Physics and Applied Mathematics Unit, Indian Statistical Institute, Kolkata-700108, India}

\date{\today}

\begin{abstract}
We obtain exact solutions of the (2+1) dimensional Dirac oscillator in a homogeneous  magnetic field within a minimal length ($\Delta x_0=\hbar \sqrt{\beta}$), or generalized uncertainty principle (GUP) scenario.  This system in ordinary quantum mechanics has a single left-right chiral  quantum phase transition (QPT). 
We show that a non zero minimal length turns on  a infinite number of quantum phase transitions which accumulate towards the known QPT when $\beta \to 0$.  It is also shown that the presence of the minimal length modifies the degeneracy of the states and that in this case there exist a new class of states which do not survive in the ordinary quantum mechanics limit $\beta \to 0$. 
\end{abstract}

\maketitle

\section{Introduction}
\label{intro}
The existence of an absolute minimal length is a generic prediction of string theory~\cite{GARAY:1995aa}, quantum gravity~\cite{Gross:1988aa} and black hole physics~\cite{Maggiore:1993rv}. As is well known this has 
generated a considerable amount of literature investigating quantum mechanical physical systems under a generalised uncertainty principle (GUP) scenario~\cite{Kempf:1994aa,Kempf:1995aa} with the hope to point out possible and hopefully measurable effects induced by the minimal length. For a review see for example~\cite{Hossenfelder:2012jw,Sprenger:2012uc}.
 
 In particular the Dirac oscillator model, one of the few relativistic systems exactly solvable~\cite{Cook:1971aa,Moshinsky:1989aa,Benitez:1990aa,Rozmej:1999aa,Mirza:2004aa,Sadurni:2010fk,Boumali:2013aa}, has recently attracted  attention~\cite{Bermudez:2008ab,Quimbay:2013aa} also in view of the possible applications to the physics of graphene~\cite{Quimbay:2013ab}. In \cite{Quimbay:2013ab} the authors conjecture that in graphene a Dirac oscillator coupling may arise as a consequence of the effective internal magnetic field generated by the motion of the charge carriers in the planar hexagonal lattice of the carbon atoms. This certainly warrants the interest in exact solutions of this system and further investigating  beyond the standard model scenarios such as the one taken up here i.e. the  generalised uncertainty principle.

 In a recent paper~\cite{Menculini:2013aa} we have shown that the $(2+1)$ dimensional Dirac equation in the presence of a homogeneous magnetic field $B_0$ and a minimal length ($\Delta x_0=\hbar \sqrt{\beta}$) is exactly solvable and the solutions were obtained explicitly. It was also shown~\cite{Menculini:2013aa} that the presence of a minimal length modifies the spectrum and there are states which do not have any "classical" counterparts ($\beta \to 0$ limit). Here we shall consider the same model as in \cite{Menculini:2013aa} but in the presence of a Dirac oscillator interaction. This model also turns out to be exactly solvable and we find explicit analytic solutions discussing in detail the degeneracy with respect to the angular momentum quantum number. In this context it may be recalled that this system in ordinary quantum mechanics (no minimal length or $\beta \to 0$ limit) is characterized by a left-right chiral quantum phase transition at a given critical value ($B_{\text{cr}}$) of the external magnetic field \cite{Bermudez:2007ab,Quimbay:2013aa}. Such a phase transition is characterized, apart from other things, by the spectrum which is different for $B\gtrless B_{\text{cr}}$. In particular, the spectrum is non analytic at the critical point $B=B_{\text{cr}}$. It will be seen here that this scenario is greatly altered in the presence of a minimal length. Most interestingly we find that for any non vanishing minimal length there appear an infinite multitude of quantum phase transitions which have an accumulation point at precisely the single QPT of the vanishing minimal length limit. In any neighborhood of this QPT there appear infinitely many QPTs which depend on the minimal length parameter ($\beta$).  In other words the minimal length induces many other critical magnetic fields which have the peculiarity of becoming infinitely large in the limit $\b\ra 0$.  
 
 Two of the present authors have recently discussed the Dirac oscillator in a homogenous magnetic field in the presence of non commutativity~\cite{Panella:2014aa} finding similarly that the parameter associated to the non commutativity of the momentum operators shifts the known QPT (the shift  vanishing in the commutative limit) and the parameter associated to non commutativity of the coordinates generates a second new QPT which becomes inaccessible  $B_{\text{cr}}\to\infty$ in the commuting limit.   

We would also like to remind here that very recently the one dimensional version of the Dirac oscillator has been realized experimentally~\cite{Franco-Villafane:2013aa}. Practicable prospects of realizing soon the two dimensional version of the Dirac oscillator have also been discussed~\cite{Franco-Villafane:2013aa,Sadurni:2010fk,Hul:2005uq}.  

Finally we  comment briefly on possible applications to the physics of (two dimensional) materials like graphene~\cite{Geim:2007aa,Castro-Neto:2009aa,Semenoff:1984aa}, silicene~\cite{Lalmi:2010kx,Vogt:2012vn,Fleurence:2012ys,Lin:2012zr}, and the newly discovered germanene~\cite{Davila:2014aa,Cahangirov:2009aa}, that have captured the interest of a wide scientific community, and which all share the feature that the motion of the charge carriers is known to be  described by an effective (2+1) dimensional Dirac equation.

The organisation of the rest of the paper is as follows: in section \ref{exact-sol} we formulate the problem and present exact solutions; in section \ref{qpt} we analyse the spectrum and explain $\b$ dependent quantum phase transitions and finally section \ref{con} is devoted to a conclusion.


\section{Exact solutions of the (2+1) dimensional Dirac Oscillator in a magnetic field within a minimal length}
\label{exact-sol}
In the minimal length formalism the commutation relation between  position and momentum is~\cite{Kempf:1994aa,Kempf:1995aa}:
\beq
[{\hat x}_i,{\hat p}_j]=i\hbar \d_{ij}(1+\b {\bm p}^2)\,, \label{rel1}
\eeq
where $\b>0$ is the minimal length parameter. The corresponding GUP reads:
\beq
\Delta x_i \Delta p_j\geq \f{\hbar}{2}\d_{ij}[1+\b(\Delta{\mbox{\bf p}})^2+\b\langle{\mbox{\bf p}}\rangle^2]\, ,
\eeq
and the associated minimal observable length is 
$\Delta x_0=\hbar\sqrt{\beta}$.
Among the various representations of ${\hat x}_i$ and ${\hat p}_i$ which satisfies Eq.(\ref{rel1}) we consider the following :
\beq
{\hat x}_i=i\hbar (1+\b {\bm p}^2)\frac{\partial}{\partial p_i},~~~~~{\hat p_i}=p_i\, ,\label{rep}
\eeq

The Hamiltonian of the $(2+1)$ dimensional Dirac oscillator in the presence of a homogeneous magnetic field $\bm{B}= (0,0,B_0) $ is given by 
\beq\label{h1}
H=c{\bm\s}\cdot({\hat{{\bm p}}-iM\omega\s_z\hat{\bm r}+\frac{e}{c}\hat{\bm A}})+\s_z Mc^2\, ,
\eeq
where $\bm{\s}=(\s_x,\s_y)$, and $\s_z$ are Pauli matrices and the vector potential is given by $\hat{\bm A}= (-B_0/2,B_0/2,0)$.
\begin{table*}[t!]
\begin{ruledtabular}
\begin{tabular}{cccc} 
$ \quad $ & $\qquad \qquad m$ & ${ \varphi^{(1)}_{n,m}}$ & ${ k^2}$  \\ 
\hline 
$ (a) $ & $\left\{ m \geq 0 \right\} \cap \left\{ m>-\frac{3}{2}-\frac{1}{\beta\lam}\right\}$ & $s^{\zeta_1}c^{\xi_1}\!\!\phantom{F}_2F_1\left(-n,n+\zeta_1+\xi_1,\zeta_1+\frac{1}{2};s^2\right)$ & $\frac{1}{4}\left(2n+\zeta_1+\xi_1\right)^2$  \\
\hline  
$ (b) $ & $\left\{m \leq -1\right\} \cap \left\{ m<-\frac{1}{2}-\frac{1}{\b\lam}\right\}$ & $s^{1-\zeta_1}c^{1-\xi_1}\!\!\phantom{F}_2F_1\left(-n,n+2-\zeta_1-\xi_1,\frac{3}{2}-\zeta_1;s^2\right)$& $\frac{1}{4}\left(2n+2-\zeta_1-\xi_1\right)^2$ \\
\hline
$ (c) $ & $\left\{-\frac{3}{2}-\frac{1}{\b\lam}<{m}\leq -1\right\}$ & $s^{1-\zeta_1}c^{\xi_1}\!\!\phantom{F}_2F_1\left(-n,n+1-\zeta_1+\xi_1,\frac{3}{2}-\zeta_1;s^2\right)$& $\frac{1}{4}\left(2n+1-\zeta_1+\xi_1\right)^2$ \\
\hline
$ (d) $ & $\left\{0\leq m <-\f{1}{2}-\f{1}{\b\lam}\right\}$ & $s^{\zeta_1} c^{1-\xi_1}\!\!\phantom{F}_2F_1\left(-n,n+1+\zeta_1-\xi_1,\zeta_1+\frac{1}{2};s^2\right)$& $\frac{1}{4}\left(2n+1+\zeta_1-\xi_1\right)^2$ \\
\end{tabular}
 \end{ruledtabular}
\caption{$\varphi^{(1)}_{n,m}$ and the corresponding energy values for different values of $m$. In this case a solution to Eq.~\eqref{parmunu} is $\mu=m+1+\frac{1}{\beta\lambda}$ and $\nu=m$. Note that the two classes of solutions (a) and (c) were both admissible at $m=0$, but since they are equal there, we have let the range of (c) start from $m=-1$. For brevity we have indicated $s=\sin q$ and $c=\cos q$.
} 
\label{tab:up}
\end{table*}
\begin{table*}[t]
\begin{ruledtabular}
\begin{tabular}{cccc} 
$\quad$ & $\qquad \qquad m$ & ${\varphi^{(2)}_{n,m}}$ & ${ k^2}$ \\ 
\hline 
$(a)$ &$\left\{ m \geq 0\right\} \cap \left\{m>-\f{1}{2}-\f{1}{\beta\lam}\right\}$ & $s^{\zeta_2}c^{\xi_2}\!\!\phantom{F}_2F_1\left(-n,n+\zeta_2+\xi_2,\zeta_2+\frac{1}{2};s^2\right)$ & $\frac{1}{4}\left(2n+\zeta_2+\xi_2\right)^2$  \\
\hline  
$(b)$&$\left\{m\leq-1\right\} \cap \left\{{m}<\frac{1}{2}-\frac{1}{\b\lam}\right\}$ & $s^{1-\zeta_2}c^{1-\xi_2}\!\!\phantom{F}_2F_1\left(-n,n+2-\zeta_2-\xi_2,\frac{3}{2}-\zeta_2;s^2\right)$& $\frac{1}{4}\left(2n+2-\zeta_2-\xi_2\right)^2$ \\
\hline
$(c)$&$\left\{-\frac{1}{2}-\frac{1}{\b\lam}<{ m}\leq -1\right\}$ & $s^{1-\zeta_2}c^{\xi_2}\!\!\phantom{F}_2F_1\left(-n,n+1-\zeta_2+\xi_2,\frac{3}{2}-\zeta_2;s^2\right)$& $\frac{1}{4}\left(2n+1-\zeta_2+\xi_2\right)^2$ \\
\hline
$(d)$&$\left\{ -1 \leq { m}<\frac{1}{2}-\frac{1}{\b\lam}\right\}$ & $s^{\zeta_2}c^{1-\xi_2}\!\!\phantom{F}_2F_1\left(-n,n+1+\zeta_2-\xi_2,\zeta_2+\f{1}{2};s^2\right)$& $\frac{1}{4}\left(2n+1+\zeta_2-\xi_2\right)^2$ \\
\end{tabular}
\end{ruledtabular}
\caption{$\varphi^{(2)}_{n,m}$ and the corresponding energy values for different values of $m$. In this case a solution to Eq.~\eqref{parmunu} is $\mu=m+\frac{1}{\beta\lambda}$ and $\nu=m+1$. In writing the constraints for the first and third row, use has been made of the fact that for $m=-1$ the two solutions and the spectrum are exactly the same; thus we have arbitrarily chosen to let the case (a) start from $m=0$ while including the case $m=-1$ in (c).
Here again $c=\cos q$, $s=\sin q$.} \label{tab:down}
\end{table*}
The eigenvalue problem corresponding to Eq.(\ref{h1}) reads
\beq H \psi =
\left(\ba{cc}Mc^2 & cP_-\\cP_+ & -Mc^2\ea\right)\left(\ba{c}\psi^{(1)} \\ \psi^{(2)}\ea\right)=E\left(\ba{c}\psi^{(1)} \\ \psi^{(2)}\ea\right)\,. 
\label{H}
\eeq
Introducing polar coordinates in the momentum space $(p,\vart)=(\sqrt{p_x^2+p_y^2}, \tan^{-1}\displaystyle\frac{p_y}{p_x})$ the operators $P_{\pm}$ read:
\begin{equation}
P_{\pm}= e^{\pm i\vart}\left[p\mp\lam\left(1+\b p^2\right)\left(\partial_p\pm\frac{i}{p}\partial_\vart\right) \right] \, ,
\label{polPmETA}
\end{equation}
where 
$\lam=\f{\hbar eB_0}{2c}-M\hbar\omega=M\hbar \left({\widetilde{\omega}_c}-\omega\right)$ and $\widetilde{\omega}_{c}=\f{eB_0}{2Mc}$.
We observe that depending on the strength of the magnetic field (in comparison to the oscillator strength), $\lam$ can be either positive or negative.
In order to solve the eigenvalue equations of $\psi^{(1,2)}$ we make the ansatz:
\beq
\ba{l}
\psi^{(1)}=e^{im\vart}p^{-1/2}\varphi^{(1)}_m(p)\,,~~~\psi^{(2)}=e^{i(m+1)}p^{-1/2}\varphi^{(2)}_m(p)\, \\
p=\f{1}{\sqrt{\b}}\tan(q), \qquad q=x/2+\pi/4
\ea
\eeq
and obtain the second-order equations ($i=1,2$):
\beq \label{xeqn}
\left\lbrace-\diff[2]{}{x}+V_i(x)\right\rbrace\varphi^{(i)}_m(x)=k^2\varphi^{(i)}_m(x),
\eeq 
where $k^2=\frac{\eps^2+1/\b}{4\b\lambda^2}$ with $\epsilon^2=\epsilon_+\,\epsilon_-$, $\epsilon_\pm= (E\pm Mc^2)/c$ and:
\beq
\label{potential(x)}
V_i(x)=\left(\f{\mu_i^2+\nu_i^2}{2}-\f{1}{4}\right)\f{1}{\cos^2x}+\f{\mu_i^2-\nu_i^2}{2}\f{\sin x}{\cos^2x}
\eeq
\begin{equation}
\label{parmunu}
\mu_i=\xi_i-\frac{1}{2} \, , \qquad  \nu_i=\zeta_i-\frac{1}{2}\, ,
\end{equation}
\beq
\zeta_i=m-\frac{1}{2}+i ,~~\xi_i=m+\frac{5}{2}-i+\frac{1}{\b\lam}\,.\label{xizeta2}
\eeq


The physically acceptable solutions to the above equations can be obtained from \cite{Levai:2006aa} as (omitting the index $i$)
\begin{eqnarray}
\label{levair}
\psi_n(x)&=&C\,\,\displaystyle[z(x)]^{\f{\mu}{2}+\f{1}{4}}[1-z(x)]^{\f{\nu}{2}+\f{1}{4}}\nonumber\\
&&\phantom{xxxxxx}\,\!\!\phantom{F}_2F_1\left(-n,\mu+\nu+1;\nu+1;1-z(x)\right)\nonumber\\
k_n&=&n+\displaystyle\frac{\mu+\nu+1}{2}
\end{eqnarray}
where $z(x)=\displaystyle\f{1-\sin x}{2}=\cos^2(q)$ and $C$ is a normalization constant. 

The vanishing of the wave-function at the end-points (i.e. $q=0$ and $q=\pi/2$) is ensured by enforcing the following constraints (remembering the symmetry of $V(x)$ through the replacements $\mu_i \to -\mu_i$ and/or $\nu_i \to -\nu_i $): 
(a) $\mu_i>-1/2$ and $\nu_i > -1/2$;
(b) $\mu_i<1/2$ and $\nu_i < 1/2$;
(c) $\mu_i >  -1/2 \text{\ and\ }\nu_i< 1/2 $;
(d) $\mu_i <  1/2 \text{\ and\ }\nu_i>- 1/2 $.
Solving for the parameters $\mu_i ,\nu_i$ in Eqs.~(\ref{parmunu}) in terms of the angular momentum quantum number $m$ provides four ranges (of $m$) {and the wave functions (from Eq.~\eqref{levair})} in Tables \ref{tab:up} and \ref{tab:down}.


\begin{table*}[t]
\begin{ruledtabular}
\begin{tabular}{ccc} 
range  & $ \psi^{(1)} $ class of solution & $\psi^{(2)}$ class of solution\\
\hline 
{$\rho>\rho^*$}& (a)  $m\geq 0$ ;~~  (c)  $m=-1$ ;~~   (b)  $m\leq -1$ & (a)  $m\geq 0$; ~~ (d)  $m=0, \mathbf{-1}$;~~   (b)  $m\leq \mathbf{-1}$ \\
\hline  
{$0<\rho<\rho^*$}& (a)  $m\geq 0$;~~~ (c)  $\tau_-<m\leq -1$;~~~(b)  $m<\tau_+$   &(a)  $m\geq 0$;~~~ (c)  $\tau <m\leq -1$;~~~(b)  $m<\tau_+$\\
\hline
{$-\rho^*<\rho<0$}& (a)  $m\geq 0 \wedge m>\tau_-$;~~(d) $0\leq m < \tau$;~~(b)  $m\leq -1$& (a) $m>\tau>0$;~~(d)  $\mathbf{-1}\leq m < \tau_+$;~~(b)  $m\leq \mathbf{-1}$ \\
\hline
{$\rho<-\rho^*$}& (a)  $m\geq 0$;~~(c)  $ m = -1$;~~(b) $m\leq -1$ & (a)  $m\geq0$;~~
 (d)  $m =0,\mathbf{-1}$;~~ (b)  $m\leq \mathbf{-1}$ \\
\end{tabular}
\end{ruledtabular}
\caption{Values of $m$ which are repeated in  boldface in two different classes of solutions within the same range of $\rho$ are associated to wave functions and spectrums which coincide, so that only one need be considered, e.g. we omit the case $m=-1$ for the solution (d) in the first and last range  
($\tau=-\f{1}{2}-\f{\rho^*}{2\rho}$ and $\tau_\pm = \tau \pm 1$).}
\label{tab:division}
\end{table*}
To understand the physical meaning of the quantity $\f{1}{\b\lam}$ we express it as
\beq
\f{1}{\b\lam}=\f{\hbar}{(\hbar\sqrt{\beta})^2 M\left({\widetilde{\omega}_c}-\omega\right)}=\left[(\Delta x_0)^2 \left(\f{1}{\ell_L^2}-\f{1}{\ell_D^2}\right)\right]^{-1}
\eeq
where $\ell_L$, $\ell_D$ are the characteristic lengths of the associated Landau, $\ell_L=\sqrt{\f{\hbar}{M\widetilde{\omega}_c}}$, and Dirac, $\ell_D=\sqrt{\f{\hbar}{M\omega}}$, oscillators. The quantity in brackets is certainly a small one when the two lengths are very different, because in this case the difference is dominated by one of the two terms, and we assume the condition
$\Delta x_0 \ll \ell_L\,,\, \ell_D$ 
to be always satisfied. Furthermore, the same quantity in brackets is also obviously small when the two lengths are equal or nearly so $\ell_L^2 \simeq \ell_D^2$, so in both these situations we have $\f{1}{\lam\beta} \gg 1$. However, we now expect there will be  certain values of $\ell_{L,D}$ with: 
$(\Delta x_0)^2 \left(\f{1}{\ell_L^2}-\f{1}{\ell_D^2}\right) \simeq 1$
so that $\f{1}{\lam\beta}$ will be of order unity. 
From now on, we work with the dimensionless parameter $\rho$, defined as
\beq
\label{defrho}
\rho=\frac{\hbar\left({{\widetilde{\omega}}_c}-\omega\right)}{M c^2}
\eeq
so that $\lam=\rho M^2 c^2$.
 To construct the full spinor solutions, we proceed by first studying how the decoupled solutions appear in the possible ranges of the parameter $\rho$. We first identify these in Table~\ref{tab:division}.

Now, care has to be taken in building up the spinors. Obviously, we can put together to form a bispinor only those eigenfunctions $\psi^{(1)}$ and $\psi^{(2)}$ that have the same energy $E$ and are paired through the intertwining relations from Eq.~\eqref{H}. We shall write down in tables \ref{tab:range1}, \ref{tab:range2} and \ref{tab:range3} the complete solutions with their respective energy eigenvalues, explicitly in terms of a unique normalization constant $C$ of the Dirac spinor.
Defining a new constant $\rho^*=\frac{2}{\b M^2 c^2} = 2\, \widetilde\lambda^2_{\text{C}}/(\hbar\sqrt{\beta})^2$,  with $\widetilde\lambda_C=\hbar/(Mc)$ the particle's reduced Compton wavelength, the two values
$\rho_{\pm}=\pm \rho^*$ 
define the various ranges.
\begin{table*}[h]
\caption{\label{tab:range1}Energy levels and the corresponding wave functions constructed from the first (and last) row of Table (\ref{tab:division}). Note that the solutions of classes (c) and (d), for the up and down component respectively, cannot be used to form a spinorial solution and must be discarded. ($\rho>+\rho^{*}\, \vee 	\, \rho<-\rho^{*}$)}
\begin{ruledtabular}
\begin{tabular}{cc} 
{$m\geq 0$}& $\f{E}{Mc^2}=\pm\sqrt{1+4\rho(n+m+1)\left[1+\frac{2\rho}{\rho^*}(n+m+1)\right]}\qquad n=0,1,\ldots$\qquad
$\psi_{n,m} =C
\begin{pmatrix} 
\psi_{n,m}^{(1)} \\
\f{\eps_{-}}{2\rho M^2 c^2(m+1)}\psi_{n,m}^{(2)} 
\end{pmatrix} $ \\ 
&$\psi_{n,m}^{(1)}=\frac{p^{m}\!\!\phantom{F}_2F_1\left(-n,n+2(m+1)+\frac{\rho^*}{2\rho},m+1,\frac{\b p^2}{1+\b p^2}\right)}{\left(1+\b p^2\right)^{m+1+\frac{\rho^*}{4\rho}}}\,e^{im\vart}$ \qquad
$ \psi^{(2)}_{n,m}=\frac{p^{(m+1)}\!\!\phantom{F}_2F_1\left(-n,n+2(m+1)+\frac{\rho^*}{2\rho},m+2,\frac{\b p^2}{1+\b p^2}\right)}{\left(1+\b p^2\right)^{m+1+\frac{\rho^*}{4\rho}}}\,e^{i(m+1)\vart}$\\
\hline
{$m\leq -1$} & $\f{E}{Mc^2}=\pm\sqrt{1+4\rho(n+|m|)\left[\frac{2\rho}{\rho^*}(n+|m|)-1\right]}\qquad n=0,1,\ldots$ \qquad
$\psi_{n,m} =C
\begin{pmatrix} 
\psi_{n,m}^{(1)} \\
\f{2\rho M^2 c^2 m}{\eps_{+}}\psi_{n,m}^{(2)} 
\end{pmatrix} $\\
&$\psi_{n,m}^{(1)} =\frac{p^{|m|}\!\!\phantom{F}_2F_1\left(-n,n+2|m|-\frac{\rho^*}{2\rho},1+\left|m\right|,\frac{\b p^2}{1+\b p^2}\right)}{\left(1+\b p^2\right)^{|m|-\frac{\rho^*}{4\rho}}}\,e^{im\vart}$\qquad
$\psi_{n,m}^{(2)} =\frac{p^{(|m|-1)}\!\!\phantom{F}_2F_1\left(-n,n+2|m|-\frac{\rho^*}{2\rho},|m|,\frac{\b p^2}{1+\b p^2}\right)}{\left(1+\b p^2\right)^{|m|-								\frac{\rho^*}{4\rho}}}\,e^{i(m+1)\vart}$ \\ 
\end{tabular} 
\end{ruledtabular}
\end{table*}
\begin{table*}[!]
\caption{\label{tab:positive}Energy levels and the corresponding wave functions constructed from the second row of Table (\ref{tab:division}). Note that the ground state in the negative branch is a singlet state with negative energy $E=-Mc^2$. ($0<\rho<+\rho^{*}$)}
\begin{ruledtabular}
\begin{tabular}{cc} 
\multirow{2}{*}{$m\geq 0$}& $\f{E}{Mc^2}=\pm\sqrt{1+4\rho(n+m+1)\left[1+\frac{2\rho}{\rho^*}(n+m+1)\right]}\qquad n=0,1,\ldots$\qquad
$\psi_{n,m} =C
\begin{pmatrix} 
\psi_{n,m}^{(1)} \\
\f{\eps_{-}}{2\rho M^2 c^2(m+1)}\psi_{n,m}^{(2)} 
\end{pmatrix} $ \\ 
&$\psi_{n,m}^{(1)}=\frac{p^{m}\!\!\phantom{F}_2F_1\left(-n,n+2(m+1)+\frac{\rho^*}{2\rho},m+1,\frac{\b p^2}{1+\b p^2}\right)}{\left(1+\b p^2\right)^{m+1+\frac{\rho^*}{4\rho}}}e^{im\vart}$ \qquad
$ \psi^{(2)}_{n,m}=\frac{p^{(m+1)}\!\!\phantom{F}_2F_1\left(-n,n+2(m+1)+\frac{\rho^*}{2\rho},m+2,\frac{\b p^2}{1+\b p^2}\right)}{\left(1+\b p^2\right)^{m+1+\frac{\rho^*}{4\rho}}}e^{i(m+1)\vart}$\\
\hline
{$\tau<m\leq -1$}& $\f{E}{Mc^2}=-1\,,\,\,\psi_{0,m}=C\begin{pmatrix}
0 \\
\psi_{0,m}^{(2)} 
\end{pmatrix}$;~~
$ \f{E}{Mc^2}=\pm\sqrt{1+4\rho n \left[1+\frac{2\rho}{\rho^*} n\right]},\quad n=1,2,\ldots  \,\,\, \psi_{n,m}=C\begin{pmatrix}
\psi_{n-1,m}^{(1)} \\
\f{2m\rho M^2 c^2}{\eps_{+}}\psi_{n,m}^{(2)} 
\end{pmatrix}$ \\ 
&{$ \psi_{n,m}^{(1)} =\frac{p^{|m|}\!\!\phantom{F}_2F_1\left(-n,n+2+\frac{\rho^*}{2\rho},|m|+1,\frac{\b p^2}{1+\b p^2}\right)}{\left(1+\b p^2\right)^{1+\frac{\rho^*}{4\rho}}}e^{im\vart}$}\qquad
{$ \psi_{n,m}^{(2)} =\frac{p^{|m+1|}\!\!\phantom{F}_2F_1\left(-n,n+\frac{\rho^*}{2\rho},|m|,\frac{\b p^2}{1+\b p^2}\right)}{\left(1+\b p^2\right)^{\frac{\rho^*}{4\rho}}}e^{i(m+1)\vart}$}  \\
\hline
{$m< \tau$} & $\f{E}{Mc^2}=\pm\sqrt{1+4\rho(n+|m|)\left[\frac{2\rho}{\rho^*}(n+|m|)-1\right]}\qquad n=0,1,\ldots$ \qquad 
$\psi_{n,m} =C
\begin{pmatrix} 
\psi_{n,m}^{(1)} \\
\f{2\rho M^2 c^2 m}{\eps_{+}}\psi_{n,m}^{(2)} 
\end{pmatrix} $\\
&$\psi_{n,m}^{(1)} =\frac{p^{|m|}\!\!\phantom{F}_2F_1\left(-n,n+2|m|-\frac{\rho^*}{2\rho},1+\left|m\right|,\frac{\b p^2}{1+\b p^2}\right)}{\left(1+\b p^2\right)^{|m|-\frac{\rho^*}{4\rho}}}e^{im\vart}$\qquad
$\psi_{n,m}^{(2)} =\frac{p^{(|m|-1)}\!\!\phantom{F}_2F_1\left(-n,n+2|m|-\frac{\rho^*}{2\rho},|m|,\frac{\b p^2}{1+\b p^2}\right)}{\left(1+\b p^2\right)^{|m|-								\frac{\rho^*}{4\rho}}}e^{i(m+1)\vart}$ \\ 
\end{tabular} 
\label{tab:range2}
\end{ruledtabular}
\end{table*}

It can be readily observed that in the limit $\beta\to0$ ($\rho^*\to \infty$) the range described in Table \ref{tab:range1} ($\rho>+\rho_{*}\, \vee 	\, \rho<-\rho_{*}$) disappears, leaving only the ranges described in Tables \ref{tab:range2} ($0 < \rho < \rho^*$) and \ref{tab:range3}, ($-\rho_{*}<\rho<0$) respectively for positive and negative values of $\rho$. Thus when $\beta \to 0$ we recover the correct description of the system in terms of a unique quantum phase transition at $\rho=0$, or from Eq.~\eqref{defrho}, at the critical field:
\begin{equation}
\label{bcrit}
B_{\text{cr}}=\frac{2Mc}{e}\, \omega
\end{equation} 
through which we pass from a phase (characterized by $\rho<0$) where the ground state of the positive branch of the spectrum is given by $E=Mc^2$ to another ($\rho >0$) where the ground state is given by ${E}={Mc^2}\,\sqrt{1+4\rho
}$.\\

\section{Quantum Phase Transition(s)}
\label{qpt}
Here we highlight the new features brought about by the presence of a minimal length, i.e. by a nonzero $\b$ parameter. We discuss explicitly only  the positive branch of the spectrum but a similar analysis can be applied as well to the negative branch. As stated before, we must now handle four different regimes, two of which (the extremal ones) are absent in the standard quantum mechanical framework as their outbreak is at $\rho=\pm\rho^{*}$ and these points go respectively to $\pm \infty$ when $\b \to 0$. The spectrum in this  ``external" range, see Table~\ref{tab:range1}, is given by the levels: $
{E}={Mc^2}\sqrt{1+4\rho N[2(\rho/\rho^*) N \pm 1]}$
where the plus and minus signs refer respectively to $m\geq0$ (where $N=n+m+1$) and to $m\leq-1$ (where $N=n+|m|$). In either case $N=1,2,\ldots$ and the degeneracy of each distinct level is $D=N$. Note that the ground state is represented by the eigenvalue corresponding to $N=1$ with the sign $-$ or $+$ depending on the positive or negative sign of $\rho$. \\
We now turn to the range as in Table \ref{tab:range2}. Defining
$\tau=-\f{1}{2}-\f{1}{\b\rho M^2 c^2}$
in this range we have $-\infty<\tau<-1$. Here the spectrum is interestingly given by:\\
(i) the levels ${E}={Mc^2}\sqrt{1+4\rho N[2(\rho/\rho^*) N + 1]}$, $N=1,2,\ldots$ with $N=n+m+1$ for $m\geq 0$ (degeneracy $D=N$) and $N=n$ ($n$ starting from 1) for $\tau<m\leq-1$ (degeneracy $D=|{\left \lceil{\tau}\right \rceil} |$); the total degeneracy of a level is $D=N+|{\left \lceil{\tau}\right \rceil} |$ \\
(ii) the levels ${E}={Mc^2}\sqrt{1+4\rho N[2(\rho/\rho^*) N - 1]}$ with $N=n+|m|$ for $m<\tau$ (it follows that $N=2,3,\ldots$) and a degeneracy which varies from $D=0$ when $\rho\to 0^{+}$ (the level does not exist) and then increasing gradually to reach the maximum $D=N-1$ when $\rho \to\rho_{+}$.\\
\begin{figure*}[t]
\centering
\includegraphics[width=16cm]{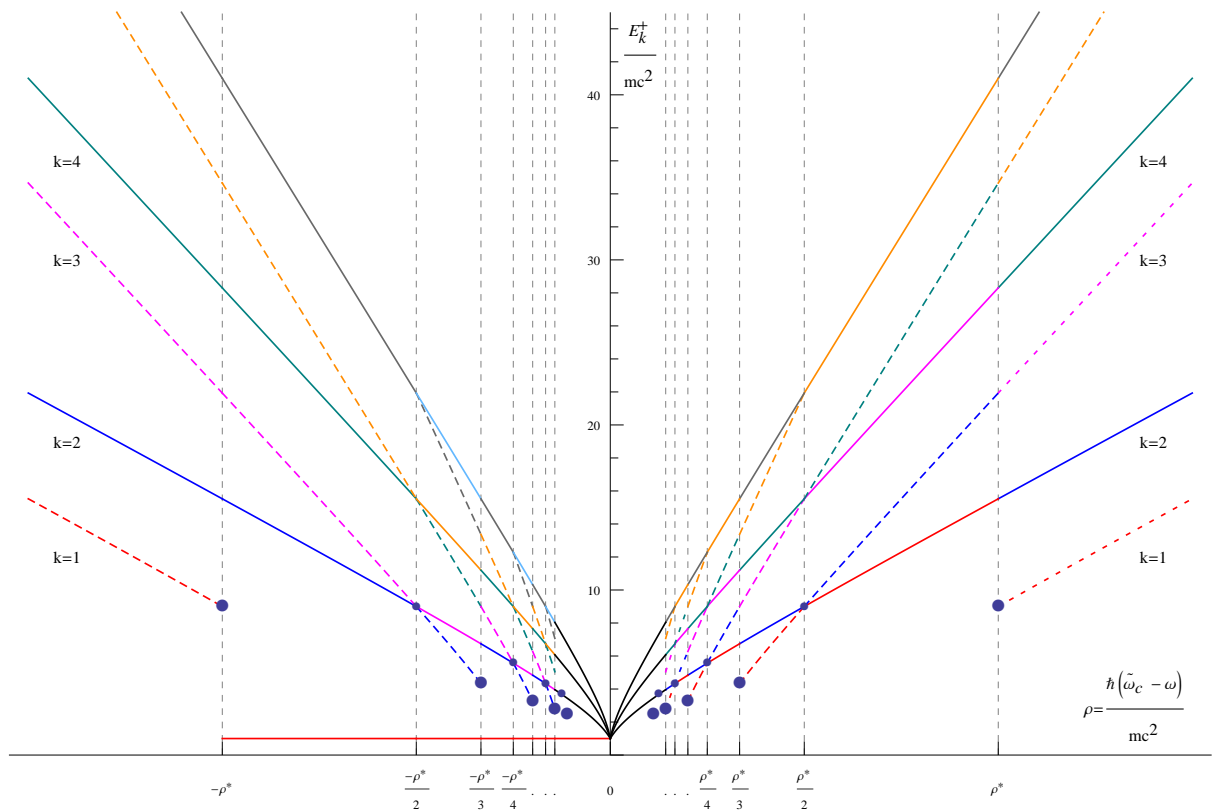}
\caption{(Color online) Plot of the first few energy levels (ordered by increasing energy at any given $\rho$, for  $\rho^*=20$. Solid lines represent states which persist in the $\beta \to 0$ limit. Dashed lines refer to states which exist only for $\beta \neq 0$, disappearing altogether when one takes the limit $\b\to 0$ (their threshold $\rho_{\cal N}=\rho^*/{\cal N}, {\cal N} = 1, 3, \dots$  becoming increasingly large since $\rho^* \to \infty$).  Each energy level is represented by a broken line alternatively solid and dashed (of the same color). Points of discontinuities and change of slope in the energy levels signal a quantum phase transition. Each energy level crosses the critical point at $\rho=0$ undergoing infinitely many quantum phase transitions. }
\label{fig:spectrum}
\end{figure*}
\begin{table*}[h]
\caption{\label{tab:negative}Energy levels and the corresponding wave functions constructed from the third row of Table (\ref{tab:division}). Note that here the ground state in the positive branch is a singlet with positive energy $E=Mc^2$. ($-\rho^{*}<\rho<0$)}
\begin{ruledtabular}
\begin{tabular}{cc} 
{$m>\tau$}& $\f{E}{Mc^2}=\pm\sqrt{1+4\rho(n+m+1)\left[1+\frac{2\rho}{\rho^*}(n+m+1)\right]}\qquad n=0,1,\ldots$\qquad
$\psi_{n,m} =C
\begin{pmatrix} 
\psi_{n,m}^{(1)} \\
\f{\eps_{-}}{2\rho M^2 c^2(m+1)}\psi_{n,m}^{(2)} 
\end{pmatrix} $ \\ 
&$\psi_{n,m}^{(1)}=\frac{p^{m}\!\!\phantom{F}_2F_1\left(-n,n+2(m+1)+\frac{\rho^*}{2\rho},m+1,\frac{\b p^2}{1+\b p^2}\right)}{\left(1+\b p^2\right)^{m+1+\frac{\rho^*}{4\rho}}}e^{im\vart}$ \qquad
$ \psi^{(2)}_{n,m}=\frac{p^{(m+1)}\!\!\phantom{F}_2F_1\left(-n,n+2(m+1)+\frac{\rho^*}{2\rho},m+2,\frac{\b p^2}{1+\b p^2}\right)}{\left(1+\b p^2\right)^{m+1+\frac{\rho^*}{4\rho}}}e^{i(m+1)\vart}$\\
\hline
{$0\leq m <\tau$}& $\f{E}{Mc^2}=1 $,~~$n=0, \,\,\,\psi_{0,m}=C\begin{pmatrix}
\psi_{0,m}^{(1)}\\ 0
\end{pmatrix}$;~~$ \f{E}{Mc^2}=\pm\sqrt{1+4\rho n\left[\frac{2\rho}{\rho^*} n-1\right]} $,\quad $n=1,2,\ldots \,\,\, \psi_{n,m}=C\begin{pmatrix}
\psi_{n,m}^{(1)} \\
\f{2m\rho M^2 c^2}{\eps_{+}}\psi_{n-1,m}^{(2)} 
\end{pmatrix}$ \\ 
&{$ \psi_{n,m}^{(1)} =\frac{p^{m}\!\!\phantom{F}_2F_1\left(-n,n-\frac{\rho^*}{2\rho},m+1,\frac{\b p^2}{1+\b p^2}\right)}{\left(1+\b p^2\right)^{-\frac{\rho^*}{4\rho}}}\,\,e^{im\vart}$} \qquad
{$ \psi_{n,m}^{(2)} =\frac{p^{m+1}\!\!\phantom{F}_2F_1\left(-n,n+2-\frac{\rho^*}{2\rho},m+2,\frac{\b p^2}{1+\b p^2}\right)}{\left(1+\b p^2\right)^{1-\frac{\rho^*}{4\rho}}}\,\,e^{i(m+1)\vart}$}  \\
\hline
{$m\leq -1$} & $\f{E}{Mc^2}=\pm\sqrt{1+4\rho(n+|m|)\left[\frac{2\rho}{\rho^*}(n+|m|)-1\right]}\qquad n=0,1,\ldots$ \qquad 
$\psi_{n,m} =C
\begin{pmatrix} 
\psi_{n,m}^{(1)} \\
\f{2\rho M^2 c^2 m}{\eps_{+}}\psi_{n,m}^{(2)} 
\end{pmatrix} $\\
&$\psi_{n,m}^{(1)} =\frac{p^{|m|}\!\!\phantom{F}_2F_1\left(-n,n+2|m|-\frac{\rho^*}{2\rho},1+\left|m\right|,\frac{\b p^2}{1+\b p^2}\right)}{\left(1+\b p^2\right)^{|m|-\frac{\rho^*}{4\rho}}}e^{im\vart}$\qquad
$\psi_{n,m}^{(2)} =\frac{p^{(|m|-1)}\!\!\phantom{F}_2F_1\left(-n,n+2|m|-\frac{\rho^*}{2\rho},|m|,\frac{\b p^2}{1+\b p^2}\right)}{\left(1+\b p^2\right)^{|m|-								\frac{\rho^*}{4\rho}}}e^{i(m+1)\vart}$ \\ 
\end{tabular} 
\label{tab:range3}
\end{ruledtabular}
\end{table*}
What happens is that when $\rho$ is very close to zero, only the solution (i) is meaningful, the other one being restricted to values of $m$ which become increasingly more negative. However, when $\rho$ starts increasing we see that more and more levels pertaining to (ii) pop out and when $\rho$ is near $\rho_{+}$ we are found with a swarm of energy levels which are the same as those found in rightmost range, but without what is the ground state level in that range. All this means, in other words, that in this zone we witness the appearance of the class of solutions (ii) gradually, i.e. distinctly for each (negative) value of $m$, until in passing out to the rightmost zone (i.e. for strong magnetic fields) at the point $\rho=\rho_{+}$ the last (and lower) level turn up: that is to say the level corresponding to  $N=1$ with the energy eigenvalue given by solutions of type \text{ii} (with a minus sign in the square bracket term), corresponding to $m=-1$ and $n=0$).\\
Finally, we have the case described by Table \ref{tab:range3}. The energy eigenvalues here go as follows (note that here $0<\tau<+\infty$):\\
(iii) ${E}={Mc^2}\sqrt{1+4\rho N[2(\rho/\rho^*) N + 1]}$, with $N=n+m+1$ for $m> \tau$ (so $N=2,3,\ldots$). The degeneracy of each level goes from $D=0$ (level not permissible) when $\rho\to 0^{-}$ to $D=N-1$ when $\rho \to \rho_{-}$. \\
(iv) $E=Mc^2$ for $0\leq m < \tau$ ($D=|{\left \lceil{\tau}\right \rceil} |$).\\
(v) ${E}={Mc^2}\sqrt{1+4\rho N[2(\rho/\rho^*) N - 1]}$, $N=1,2,\ldots$ with $N=n$ for $0\leq m<\tau$ (degeneracy ($D={\left \lceil{\tau}\right \rceil} $)) and $N=n+|m|$ for $m\leq -1$ ($D=N$). Globally the level has degeneracy $D=N+{\left \lceil{\tau}\right \rceil} $.
The situation is now specular to the one we have already seen in the range  $\rho\ge0$. Indeed starting from $\rho=\rho_{-}$ , where $\tau\to 0^{+}$ and we have all the solutions as in (iii) at full degeneracy, together with (iv) with degeneracy 1 (only $m=0$ is possible) and (v) with degeneracy $N+1$; with increasing $\rho$ we observe that solutions of the kind (iii) gradually disappear while the levels (iv) and (v) become more and more degenerate and remain the only possible levels when we get very close to zero. In Fig.~\ref{fig:spectrum} the lowest energy levels are shown as a function of $\rho$.

 
 It is worth pointing out that none of the dashed  states survives up to the critical point $\rho=0$. Note also that due to these states  the  energy levels  (except for the zero mode singlet for $\rho <0$) are altered with respect to the ordinary case  $\beta \to 0$ solid lines in Fig.~\ref{fig:spectrum}, and discontinuities  and changes in slope arise in every energy level passing   from the external zone to the internal (when $\rho<0$) and then from the internal to the external (when $\rho>0$).
 We find that the points in the variable $\rho$: $\rho_{\cal N}= \pm\frac{\rho^*}{\cal N}\,, ({\cal N}=1,3,5 \dots) $ --large dark disks-- where the states shown in Fig.\ref{fig:spectrum} by dashed lines start,  and those: $\rho_{\cal N}= \pm\frac{\rho^*}{{\cal N}}\,, ({\cal N}=2,4,6 \dots)$ --small dark disks--  where such states cross the ones represented by solid lines, realise a partition of the intervals $[-\rho^*,0]\, ,\,[0,+\rho^*]$ and to each one of them corresponds a critical value of the external magnetic field, obtained from Eq.~\eqref{defrho}, given by
\beq 
\label{bcritN}
B^{\cal{N}}_{\text{cr}}= B_{\text{cr}} + \f{4c}{\cal{N}\b e\hbar}\qquad {\cal N} = 1,2,3, \dots \, ,
\eeq
where $B_{\text{cr}}$ is given in Eq.~\eqref{bcrit}, and to each $B^{\cal{N}}_{\text{cr}}$ is associated a new quantum phase transition explicitly dependent on the minimal length parameter.  We note that the thresholds of the dashed lines $\rho_{\cal N}= \pm\frac{\rho^*}{\cal N}\,, ({\cal N}=1,3,5 \dots) $ -- large disks in Fig~\ref{fig:spectrum} -- are found form the conditions $m<\tau=$ (Table~\ref{tab:range2}) and $m>\tau$ (Table~\ref{tab:range3}), recalling that $\tau=-\frac{1}{2}-\frac{\rho^*}{2\rho}$. Similarly the critical points $\rho_{\cal N}= \pm\frac{\rho^*}{{\cal N}}\,, ({\cal N}=2,4,6 \dots)$ --small disks in Fig.~\ref{fig:spectrum}-- are obtained by equating the eigenvalues of the type (i)  with those of type (ii) in the discussion above. 

In the ordinary quantum mechanical limit ($\beta \to 0$) all the  critical points $\rho_{\cal N}$ move off to infinity  together with $\rho^*$ and so they will the critical magnetic fields in Eq.~\ref{bcritN} ($\lim_{\beta\to 0} B^{{\cal N}}_{\text{cr}}=\infty$) and only the $\rho=0$ critical point ($B_{\text{cr}}=\f{2Mc}{e}\om$) and its well known associated left-right chiral quantum phase transition survive.  On the contrary it is found that $\lim_{{\cal N}\to \infty} B^{\cal N}_{\text{cr}} = B_{\text{cr}}$ which expresses the fact that the critical field $B_{\text{cr}}$ within ordinary quantum mechanics (no minimal length) acts as an accumulation point for the multitude of new critical fields $B^{\cal{N}}_{\text{cr}}$ which arise with a non vanishing minimal length.  We note that a similar behaviour has been found within the non commutative Dirac oscillator~\cite{Panella:2014aa} where the parameter related to space non commutativity simply shifts the known QPT at $\rho=0$ while the momentum non commutativity parameter introduces a second QPT which however moves out to $\rho\to \infty$ (and hence disappears) when such parameter is turned off.

One might also make the following  interesting consideration. Suppose that the oscillator frequency is such that the critical field $B_{\text{cr}}$ in eq.~\eqref{bcrit} can be realized in a laboratory and that the minimal length is so small that the quantity $B^1_{\text{cr}}-B_{\text{cr}}$ is instead too large to be realized  in a laboratory. Then one might consider the quantity $B^2_{\text{cr}}-B_{\text{cr}}= \frac{1}{2}( B^1_{\text{cr}}-B_{\text{cr}})$ and if even $B^2_{\text{cr}}$ is not experimentally accessible one can try to go to $B^3_{\text{cr}}-B_{\text{cr}}=\frac{1}{3} (B^1_{\text{cr}}-B_{\text{cr}})$ and so on until for some sufficiently large ${\cal N}$ the critical field  $B^{\cal N}_{\text{cr}}$ and the associated quantum phase transition will be accessible in the laboratory.  

 Finally we comment briefly on possible applications of the results presented in this work to the physics of new materials like graphene, silicene and germanene.
The motion of the charge carriers in such materials is known to be described by an effective (2+1)-dimensional Dirac Equation.
In reference~\cite{Panella:2014aa} it was shown that in both graphene and silicene the quantum phase transitions induced by the non commutativity will affect differently the two inequivalent Dirac points $K$ and $K'$. Indeed the quantum phase transitions take place at different critical values of the magnetic field. A similar behavior is worth investigating  in the case of the present model with a minimal length.
The result presented here could be applied directly to silicene because there the charge carriers are massive. In the case of graphene however the charge carriers are massless and  the results of this work will have to be discussed in the massless limit.

\section{Conclusion}\label{con}
Here we have obtained exact solutions of the Dirac oscillator in the presence of a homogeneous magnetic field within the generalized uncertainty principle scenario. It has been shown that unlike the $\b=0$ case in the present model there are an infinite number of quantum phase transitions depending on the minimal length parameter $\b$. In other words there are an infinite number of critical magnetic fields ($B^{\cal N}_{\text{cr}},\, {\cal N} = 1,2,3,\dots$) whose magnitude depends on $\beta$. However all these critical magnetic fields become infinitely large   in the $\b\ra 0$ limit and one is left only with the "classical" $B_{\text{cr}}$~\eqref{bcrit}.  Equally interesting is the fact that however small is $\beta$,  in any neighbourhood of $\rho=0$, or $B=B_{\text{cr}}$, there will be an infinite number of critical fields $B^{\cal N}_{\text{cr}}$. 

 We have briefly commented about possible applications of our exact solutions to the physics of materials like graphene, silicene and germanene. 

While the main scope of the present work is to report the striking appearance when $\beta\ne0$ of an infinite number of quantum phase transitions in the vicinity of the $\rho=0$ critical point a detailed study of the effect of such QPTs  on the thermodynamic functions of the system will be the object of further investigations. 

\begin{acknowledgments}
 P.~R. acknowledges hospitality from the Physics Department of the
University of Perugia and financial support from INFN - Istituto Nazionale di Fisica Nucleare - Sezione di Perugia.
\end{acknowledgments}

\end{document}